\documentstyle[12pt]{article}
\begin{document}
\thispagestyle{empty}
\begin{center}
{\Large Cylindrical ideal magnetohydrodynamic equilibria\\ \vspace{3mm}
	   with incompressible flows \vspace{5mm} }\\
{\large G. N. Throumoulopoulos$^{\dag}$
 and H. Tasso$^{\star}$\\ 
\vspace{3mm}
{\sl $^{\bf {\dag}}$ Section of Theoretical Physics,\\
Physics Department, University of Ioannina, \\ 
 GR  451 10 Ioannina, Greece \\ \vspace{2mm}
  ${\bf ^{\star}}$ Max-Planck-Institut f\"{u}r Plasmaphysik,\\ \vspace{1mm} 
  EURATOM Association, D-85748 Garching, Germany} }
\end{center}
\vspace{2mm}
\vspace{0.5cm}
%
\begin{center}
{\large\bf Abstract} 
\end{center}

It is proved that (a) the 
solutions of the ideal magnetohydrodynamic equation, which 
describe the equlibrium states of a cylindrical plasma with
purely poloidal  flow and arbitrary cross sectional shape 
[G. N. Throumoulopoulos and G. Pantis,  
Plasma Phys. and Contr. Fusion {\bf 38}, 1817 (1996)] 
are also valid for incompressible
equlibrium flows with the axial velocity component 
being a free surface quantity and (b)
for the case of isothermal incompressible equilibria  the magnetic
surfaces  have necessarily circular cross section. 
\newpage
\setcounter{page}{1}
\begin{center} 
{\large \bf I.\ \ Introduction}
\end{center}

In a recent paper \cite{ThPa} it is proved that, 
if the ideal MHD stationary flows
of a cylindrical plasma with arbitrary cross sectional shape 
are purely poloidal, they must be
incompressible. This property 
simplifies considerably
the equilibrium problem, i.e. it turns out that the equlibrium 
is governed by an elliptic partial differential equation
for the poloidal magnetic flux function $\psi$
which is amenable to 
 several classes of analytic solutions. 
For an arbitrary flow, i.e. when the velocity has 
non  vanishing axial  and
poloidal components, the equilibrium 
becomes much more complicated. With the adoption
of a specific  equation of state,
e. g. isentropic magnetic surfaces \cite{MoSo},
the symmetric equilibrium states in a
two dimensional geometry are governed by a 
partial differential equation for $\psi$, which contains 
five  surface quantities  (i.e. quantities solely dependent on $\psi$),
in conjuction with a nonlinear algebraic Bernoulli
equation. The derivation of analytic solutions
of this set of equations 
is  difficult. 

In the present note we study the equlibrium of a cylindrical
plasma with  incompressible flows  
and show that
the incompressibility condition makes it possible 
to construct  analytic equilibria, 
which constitute a generalization of the ones obtained                 
in Ref. \cite{ThPa}.  This is the subject of Sec. II.
The special 
class of incompressible equilibria with isothermal magnetic
surfaces is examined in Sec. III.  
Section IV summarizes 
our conclusions.

\begin{center}
{\large\bf II.\ \ Equilibrium equations and analytic solutions}
\end{center}

The ideal MHD equilibrium states of   plasma
		     flows are governed by the following set of
equations, written in standard notations and convenient units:
\begin{equation}
{\bf\nabla} \cdot (\rho {\bf v}) = 0 
					    \label{1}
\end{equation}
\begin{equation}
\rho ({\bf v} \cdot {\bf\nabla})  {\bf v} = {\bf j}
\times {\bf B} - {\bf\nabla} P 
					    \label{2}
\end{equation}
\begin{equation}
{\bf\nabla} \times  {\bf E} = 0 
					    \label{3}
\end{equation}
\begin{equation}
{\bf\nabla}\times {\bf B} = {\bf j }
					    \label{4}
\end{equation}
\begin{equation}
{\bf\nabla} \cdot {\bf B} = 0 
					    \label{5}
\end{equation}
\begin{equation}
{\bf E} +{\bf v} \times {\bf B} = 0.
					    \label{6}
\end{equation}
The system under consideration is a
cylindrical  plasma 
with flow and arbitrary cross sectional shape.  
For this
configuration  convenient coordinates are
$\xi$, $\eta$ and $z$ with unit basis vectors $ {\bf e}_\xi$, 
${\bf e}_\eta$,  ${\bf  e}_z$, where  ${\bf  e}_z$
is parallel to the axis of symmetry and $\xi$, $\eta$
are generalized coordinates pertaining to the
poloidal cross section.
The equilibrium quantities do not depend on $z$.
The divergence
free fields, i.e. the magnetic field $\bf B$, the current density
density ${\bf j}$ and the mass flow $\rho{\bf v}$, can be
expressed in terms of the stream functions $\psi(\xi, \eta)$, 
$F(\xi, \eta)$, $B_z(\xi, \eta)$ and $v_z(\xi, \eta)$ as
\begin{equation} {\bf B} = B_z {\bf e}_z + {\bf e}_z \times
{\bf\nabla} \psi
						\label{7}
 \end{equation}
\begin{equation}
{\bf j} = \nabla^2\psi {\bf e}_z - {\bf e}_z \times
\nabla B_z 
						\label{8}
\end{equation}
and
\begin{equation}
\rho {\bf v} =\rho v_z{\bf e}_z  +  {\bf  e}_z  \times
{\bf\nabla} F.
					       \label{9}
\end{equation}
Constant $\psi$ surfaces are the magnetic surfaces.
Eqs. (\ref{1})-(\ref{6}) can be reduced  by means of 
certain integrals of the system, wich are shown to be surface quantities.
To identify two of these
quantities,  the time independent  electric field
is expressed by ${\bf E} = - {\bf \nabla} \Phi$ and the Ohm's law
(\ref{6}) is projected along ${\bf e}_z$ and $\bf B$, respectively,
yielding
\begin{equation}
{\bf e}_z \cdot\left({\bf e}_z\times \nabla F\right)\times
		\left({\bf e}_z\times \nabla \psi\right) = 0
						\label{10}
\end{equation}
and
\begin{equation}
{\bf B}\cdot\nabla \Phi=0.
						\label{11}
\end{equation}
Eqs. (\ref{10}) and (\ref{11}) imply that $F=F(\psi)$ and  
$\Phi=\Phi(\psi)$. Two additional surface quantities
are found from the component of Eq. (\ref{6})  perpendicular to
a magnetic surface:
\begin{equation}
\frac{B_z F^\prime}{\rho} - v_z = \Phi^\prime,
					      \label{12}
\end{equation}
and from the component of the momentum conservation 
equation  (\ref{2}) along ${\bf e}_z$:
\begin{equation}
B_z-F^\prime v_z\equiv X(\psi).
					      \label{13}
\end{equation}
(The prime denotes differentiation with respect to $\psi$).
Solving the set of Eqs. (\ref{12}) and (\ref{13}) for
$B_z$ and $v_z$, one obtains
\begin{equation}
B_z = \frac{X(\psi)\rho -F^\prime(\psi)\Phi^\prime(\psi)}
	   {\rho - (F^\prime(\psi))^2}
					      \label{12a}
\end{equation}
and
\begin{equation}
v_z = \frac{F^\prime(\psi) X(\psi) -\Phi^\prime(\psi)}
	   {\rho - (F^\prime(\psi))^2}.
					      \label{13a}
\end{equation}
With the aid of Eqs. (\ref{10})-(\ref{13}), the components of 
Eq. (\ref{2}) along $\bf B$ and  perpendicular to a
magnetic surface, respectively, are put in the form
\begin{equation}
{\bf B} \cdot \left[{\bf \nabla} \left(\frac{v^2}{2}
+ v_z\Phi^\prime\right)
+ \frac{\nabla P}{\rho}\right] = 0 
					       \label{14}
 \end{equation}
and
\begin{eqnarray} 
{\bf \nabla} \cdot 
\left[\left(1- \frac{(F^\prime)^2}{\rho}\right)
{\bf \nabla}\psi\right]
+ \frac{F^{\prime\prime}F^\prime |\nabla\psi|^2}{\rho}
+ \frac{B_z\nabla B_z\cdot\nabla \psi}
     {|\nabla \psi|^2}  & &  \nonumber \\
+ \rho \frac{{\bf \nabla} \psi}{|{\bf \nabla} \psi|^2 }\cdot
\left[{\bf \nabla} \left(\frac{(F^{\prime})^2 |\nabla\psi|^2}{2\rho^2}\right)
+\frac{{\bf \nabla} P}{\rho}\right]=0.                                                                                      
					   \label{15} 
\end{eqnarray}
It is pointed out here that Eqs. (\ref{14}) and (\ref{15})
are valid for any equation of state for the plasma. 

In order
to reduce further the equilibrium equations, we employ the
incompressibiliy condition            
\begin{equation}
\nabla\cdot {\bf v} = 0.
					    \label{6a}
\end{equation}
 Then Eq. (1) implies
that the density is a surface quantity, 
\begin{equation}
\rho=\rho(\psi),
					    \label{16}   
\end{equation}
and, consequently,
Eqs. (\ref{12a}) and (\ref{13a}) yield
\begin{equation}
B_z=B_z(\psi), \ \ \ v_z=v_z(\psi).  
					     \label{17} 
\end{equation}
With the use of Eqs. (\ref{16}) and (\ref{17}), Eq. (\ref{14})
can be integrated yielding an expression for the pressure, i.e.
\begin{equation} 
P = P_s
(\psi) - \frac{F'^2}{2\rho} |{\bf \nabla} \psi|^2.
					      \label{19}    
\end{equation}
We note here that, unlike in static 
equilibria, in the presence of flow  magnetic surfaces do not coincide with
isobaric surfaces  because Eq. (\ref{2})
implies that ${\bf B} \cdot {\bf \nabla} P$ in
general differs from zero.
In this respect, the term $P_s(\psi)$ is the static part of 
the pressure which
does not vanish  when $F^\prime$ is set to zero; 
Eqs. (\ref{12a}),  (\ref{13a}) and (\ref{15}) have  a singularity when  
\begin{equation}
\frac{\left(F^\prime \right)^2}{\rho}=1. 
						  \label{20}
\end{equation}
On the basis of Eq. (\ref{9}) for $\rho {\bf v}$ and the
definitions $v_{Ap}^2\equiv\frac{\textstyle |\nabla \psi|^2}
{\textstyle \rho}$ for the Alfv\'en velocity associated with the
poloidal magnetic field and the Mach number $M^2\equiv\frac{\textstyle v^2}
{\textstyle v_{Ap}^2}$, Eq. (\ref{20}) can be written as
$
M^2= 1.
$

Assuming now $\frac{\textstyle (F^\prime)^2}{\textstyle \rho}\neq 1$,
and inserting Eq. (\ref{19}) into Eq. (\ref{15}), 
the latter reduces to
the {\em elliptic} differential equation
\begin{equation} 
\left[1 - \frac{(F^\prime)^2}{\rho}\right] {\bf \nabla}^2 \psi +
\frac{F^\prime}{\rho} \left(\frac{F^\prime}{2} 
 \frac{\rho^\prime}{\rho}
-F^{\prime\prime}\right) |{\bf \nabla} \psi|^2 
  +  \left(P_s +\frac{ B_z^2}{2}\right)^\prime = 0.
						    \label{21}
\end{equation}
The absence of any hyperbolic regime in Eq. (\ref{21})
can be understood by noting
that, as is well known from the gas  dynamics, the flow must be
compressible to allow the equilibrium differential
equation to depart from ellipticity. Eq. (\ref{21}) 
{\em does not contain the axial velocity
$v_z$}  and 
 is identical to the equation governing
cylindrical equilibria with purely poloidal flow \cite{ThPa}.
With the use of the ansatz 
$
\frac{\textstyle\rho^\prime}{\textstyle\rho} = 
2 \frac{\textstyle F^{\prime\prime}}{\textstyle F^\prime}
$,
which implies that 
$\frac{\textstyle (F^\prime)^2}{\textstyle \rho}\equiv M_c^2 = 
\mbox{const.}$,
Eq. (\ref{21}) reduces to
\begin{equation} 
{\bf \nabla}^2 \psi + 
\frac{1}{1-M_c^2} \left( P_s + \frac{B_z^2}{2}\right)^\prime = 0.
						  \label{22}
\end{equation}                                                                                                                                                                        
This  is similar in form to the equation
governing static equilibria;  the only explicit reminiscence of
flow is the presence of $M_c$. 
Eq. (\ref{22}) can be linearized
 for several choices of $P_s + \frac{\textstyle B_z^2}{\textstyle 2}$
and a variety of analytic solutions of the linearized equation  can
be
derived.  
In particular, the exact  solutions for a circular
cylindrical plasma obtained in Ref.  \cite{ThPa}
are also valid for incompressible equilibrium flows
with a free axial velocity $v_z(\psi)$.

The singularity $M_c^2 = 1$ is the limit at which the confinement
can be assured by the axial current $\nabla^2\psi$ alone.
For $M^2_c>1$ the derivative of $B_z^2/2$ must partly compensate 
for the pressure gradient.

\begin{center}
{\large\bf III.\ \ Equilibria with isothermal magnetic surfaces}
\end{center}

For fusion plasmas the thermal conduction along $\bf B$ is fast 
compared to    the heat transport perpendicular
to a magnetic surface and therefore equilibria
with isothermal magnetic surfaces  are of particular interest.
The plasma is also assumed to
obey the ideal gas  law 
$P = R \rho T$.
For this kind of equilibria, Eq. (\ref{19})
implies that $|\nabla \psi|$ is a surface quantity and consequently
from Eq. (\ref{15}) it turns out that $\nabla^2\psi$ is 
a surface quantity as well. 
Thus, the  incompressible,  $T=T(\psi)$ equilibria satisfy the
set of equations
\begin{equation}
|\nabla \psi|^2 =( g(\psi))^2
					      \label{23}
\end{equation}
and
\begin{equation}
\nabla^2 \psi = f(\psi).
					      \label{24}
\end{equation}
Eqs. (\ref{23}) and (\ref{24}) imply that, on  
 a magnetic surface the modulus of the vector
 $\nabla \psi$, which is perpendicular to this (arbitrary) magnetic surface,
 and $\nabla^2\psi$,  related to the variation of $|\nabla \psi|$,  
 are constants.
 Therefore, one  could speculate that magnetic surfaces are 
 restricted to be circular. This conjecture can be proved
 as follows.

The coordinates $\xi$, $\eta$ and $z$ are specified to be
 the Cartesian coordinates $x$, $y$, $z$. With the 
 introduction of the quantities  $p=\partial \psi/\partial x$,
$q=\partial \psi/\partial y$, $r=\partial^2 \psi/\partial x^2$
and $t=\partial^2 \psi/\partial y^2$ , Eqs. (\ref{23}) and (\ref{24})
are written in the form
\begin{equation}
p^2 + q^2 = g^2
						\label{25}
\end{equation}
and
\begin{equation}
r + t = f.
						\label{26}
\end{equation}
 The set of Eqs. (\ref{25}) and (\ref{26})
 can be  integrated by applying a  procedure suggested
 by Palumbo \cite{Pa}. Accordingly, considering the 
 functions $p$ and
 $q$ which are functions of $x$ and $y$ as functions of $x$ and
 $\psi(x,y)$ 
 one has
\begin{equation}
r=\left.\frac{\partial p}{\partial x}\right|_y=
  \left.\frac{\partial p}{\partial x}
  + p \frac{\partial p}{\partial\psi}\right|_y    
						\label{27}
\end{equation}
and
\begin{equation}
t=q\frac{\partial q}{\partial \psi}.
						\label{28}
\end{equation}
 (It is noted here that a surface function 
 $\zeta=\zeta(x,y)\equiv\zeta(\psi)$ can be employed instead of
 $\psi$). 
With the aid of Eqs. (\ref{25}), (\ref{27}) and (\ref{28}), 
Eq. (\ref{26}) reduces to 
$
\left.\frac{\textstyle\partial p}{\textstyle\partial x}\right|_\psi
=f-gg^\prime
$ 
and consequently 
\begin{equation}
p = x\left(f-gg^\prime\right) + h(\psi).
						\label{30}
\end{equation}
On a magnetic surface it holds that 
$
d\psi=\frac{\textstyle\partial \psi}{\textstyle \partial x}dx 
      + \frac{\textstyle\partial \psi}{\textstyle\partial y}dy\equiv 0, 
$      
and therefore
\begin{equation}
\left(\left.\frac{dy}{dx}\right|_\psi\right)^2
= \frac{p^2}{q^2}
= \frac{\left[x\left(f-gg^\prime\right) + h\right]^2}
       {g^2-\left[x\left(f-gg^\prime\right) + h\right]^2}.
						\label{31}
\end{equation}
Introducing the new quantities $a(\psi)\equiv f-gg^\prime$,
$X\equiv ax+h$ and $Y\equiv ay$, Eq. (\ref{31}) is put in the form
\begin{equation}
\left(\frac{dY}{dX}\right)^2 = \frac{X^2}{g^2-X^2}.
						 \label{32} 
\end{equation}
 Eq. (\ref{32}) describes  a circle on the $(x,y)$ plane
 with radius $| g|$ centred at $(-h/a, 0)$.

\begin{center}
{\large\bf IV.\ \ Conclusions}
\end{center}

It was proved that the ideal MHD equilibrium states of a cylindrical
plasma with incompressible flows and arbitrary cross section  
shape satisfy an elliptic  partial
differential equation [Eq. (\ref{21})], which  
is identical to the equation
governing cylindrical equilibria with purely poloidal flow;
the axial flow velocity is a free surface quantity.
This equation permits the construction of several classes
of analytic solutions. In particular, the exact equlibrium
solutions for a circular cylindrical plasma and purely
poloidal flow \cite{ThPa} are also valid for the present
case. In addition, it was proved that the magnetic 
surfaces of isothermal incompressible equilibria 
must have circular cross section.

It is  intersting to investigate 
symmetric incompressible equlibria
 in geometries representing more realistically the  magnetic
 confinement systems, e.g.  axisymmetric and straigth helically
 symmetric configurations. In this respect it may be noted
 here that, as proved in Ref. \cite{Ta}, the special class 
 of  axially symmetric, incompressible,
$ \beta_p=1$, MHD equilibria with purely poloidal velocity
 does not exist; the only possible stationary equilibria
 of this kind are of cylindrical shape.

\begin{center}
 {\large\bf Acknowledgments}
\end{center}
This work was conducted during a visit by one of the authors 
(G.N.T.) to  Max-Planck Institute f\"ur Plasmaphysik, Garching.
The hospitality provided at the said institute is appreciated.
G.N.T. acknowledges support by EURATOM (Mobility Contract No
131-83-7 FUSC). One of the authors (H.T.) would like to thank
Prof. D. Pfirsch for a useful discussion
\end{document}